\documentclass[a4paper,11pt]{article}
\usepackage{amsmath,graphicx,color,mathrsfs,subfigure,multicol}

\def\circa#1{\,\raise.3ex\hbox{$#1$\kern-.75em\lower1ex\hbox{$\sim$}}\,}
\makeatletter

\def\art{\@ifnextchar[{\eart}{\oart}}
\def\eart[#1]#2#3#4#5#6{{\rm #2}, {\em #3  #4} {\rm (#6) #5} ({\em #1})}
\def\hepart[#1]#2{{\rm #2, \em#1}}
\newcommand{\oart}[5]{{\rm #1}, {\em #2  #3} {\rm (#5) #4}}

\newcounter{alphaequation}[equation]
\def\thealphaequation{\theequation\hbox to
0.6em{\hfil\alph{alphaequation}\hfil}}
\def\eqnsystem#1{
\def\@eqnnum{{\rm (\thealphaequation)}}
\def\@@eqncr{\let\@tempa\relax \ifcase\@eqcnt \def\@tempa{& & &} \or
  \def\@tempa{& &}\or \def\@tempa{&}\fi\@tempa
  \if@eqnsw\@eqnnum\refstepcounter{alphaequation}\fi
\global\@eqnswtrue\global\@eqcnt=0\cr}
\refstepcounter{equation} \let\@currentlabel\theequation \def\@tempb{#1}
\ifx\@tempb\empty\else\label{#1}\fi
\refstepcounter{alphaequation}
\let\@currentlabel\thealphaequation
\global\@eqnswtrue\global\@eqcnt=0 \tabskip\@centering\let\\=\@eqncr
$$\halign to \displaywidth\bgroup \@eqnsel\hskip\@centering
$\displaystyle\tabskip\z@{##}$&\global\@eqcnt\@ne
\hskip2\arraycolsep\hfil${##}$\hfil& \global\@eqcnt\tw@\hskip2\arraycolsep
$\displaystyle\tabskip\z@{##}$\hfil
\tabskip\@centering&\llap{##}\tabskip\z@\cr}
\def\endeqnsystem{\@@eqncr\egroup$$\global\@ignoretrue} \makeatother

\newcommand{\be}{\begin{equation}}
\newcommand{\ee}{\end{equation}}
\newcommand{\bea}{\begin{eqnarray}}
\newcommand{\eea}{\end{eqnarray}}

\newcommand{\bright}{\begin{flushright}}
\newcommand{\eright}{\end{flushright}}
\newcommand{\bminip}{\begin{minipage}}
\newcommand{\eminip}{\end{minipage}}
\newcommand{\bcent}{\begin{center}}
\newcommand{\ecent}{\end{center}}

\begin{document}

\date{\mbox{}}

\title{
\vspace{-2.0cm}
\vspace{2.0cm}
{\bf \huge  New symmetries in Fierz-Pauli massive gravity}
 \\[8mm]
}

\author{
Gianmassimo Tasinato$^1$, Kazuya Koyama$^1$, Gustavo Niz$^{1,2}$
%.
\\[8mm]
\normalsize\it
$^1$ Institute of Cosmology \& Gravitation, University of Portsmouth,\\
\normalsize\it Dennis Sciama Building, Portsmouth, PO1 3FX, United Kingdom.\vspace{.5cm} \\
\normalsize\it
$^2$ Departamento de F\'{\i}sica, Universidad de Guanajuato,\\
{\it DCI, Campus Le\' on, C.P. 37150, Le\' on, Guanajuato, M\' exico.}\\[.3em]
}

\maketitle

\setcounter{page}{1}
\thispagestyle{empty}

\begin{abstract}
\noindent
We expose a new symmetry for linear perturbations around a solution of non-linear Fierz-Pauli massive gravity plus a bare cosmological constant. The cosmological constant is chosen such that the background metric is flat while the St\"uckelberg fields have a non-trivial profile. 
Around  this background, at linear order
 the new symmetry reduces the propagating degrees of freedom   to those of General Relativity, namely the massless helicity 2 modes only.
We discuss the physical consequences and possible applications of these findings.  
\end{abstract}

%%%%%%%%%%%%%%%%%%%%%%% sect 1 %%%%%%%%%%%%%%%%%%%%%%%%%%
\section{Introduction}

One of the most important properties of general relativity (GR) is that it propagates two 
%massless 
helicity-2 interacting 
degrees of freedom. This property is ensured by the gauge symmetry associated with general coordinate invariance, together  with the particular structure of the Einstein-Hilbert action. The fact that long-range fifth forces, for example due to light scalars, are not observed in the Solar system, suggests that GR is the correct theory of gravity near the Earth. On the other hand, serious theoretical issues with Einstein gravity in the infrared, as the cosmological constant problem  and the need to explain the observed current acceleration of the universe, motivate to consider  alternative theories of gravity. Typically, these theories involve
modifications of GR at short and/or large distances, and provide compelling explanations for the observed acceleration of the universe. However, in most cases they  spoil the properties that characterize GR, such as coordinate invariance or the number of degrees of freedom. These additional degrees of freedom may have unacceptable pathologies  (for example, they may be ghosts), or lead to predictions in disagreement with current observations.  However, as we are going to discuss here, on some backgrounds, new gauge symmetries might emerge, which are able to  remove the new undesired degrees of freedom.

 One of the simplest modification of GR consists of  adding a mass term for the graviton to the Einstein-Hilbert (EH) action, leading to a theory of massive gravity. This possibility was proposed long time ago by Fierz and Pauli \cite{Fierz:1939ix}, who considered a mass term which is uniquely defined at linear order in perturbations around flat space. %  but has many extensions beyond it. 
  Here we consider a theory that minimally extends the Fierz-Pauli (FP) action beyond linear order in perturbations  which is given by (we set $M_p^2/2=1$):
\be\label{action}
I \,=\, I_{EH}+I_{FP}\,=\,  \,\int d^4 x \sqrt{- g}\,\left[  R -2 \Lambda\right]- \frac{m^2}{4}\,\,\int d^4 x  \sqrt{-g}\,h_{\mu \nu} h_{\rho \sigma}\,\left(g^{\mu \rho} g^{\nu \sigma}-g^{\mu \nu } g^{\rho \sigma} \right)\,.
\ee
In the previous formula $g_{\mu \nu}$ is the  dynamical metric in four dimensions, and $h_{\mu \nu}= g_{\mu \nu}-\eta_{\mu \nu}$ represents the displacement
of the dynamical metric from a fiducial, fixed metric that we choose to be the Minkowski spacetime.
Notice that we have also added a bare cosmological constant term to the usual Einstein-Hilbert piece, controlled by the parameter $\Lambda$.

In the case $\Lambda=0$, Fierz and Pauli proved that the action describing quadratic perturbations around Minkowski space contains five massless degrees of freedom: two tensor, two
vectors, and one scalar \cite{Fierz:1939ix}. Some time later, Boulware and Deser showed that an additional propagating degree of freedom, a ghost, arises when a point-like source is added to flat space, whose kinetic term is weighted by the inverse of the source mass \cite{Boulware:1973my}. In total, around a generic background this theory propagates six degrees of freedom. This is expected since general coordinate invariance is broken by the mass term. See \cite{Hinterbichler:2011tt} for a recent review. 

However, it is interesting to ask whether a background solution for the theory (\ref{action}) exist, around which a smaller number of degrees of freedom propagate due to the emergence of new gauge symmetries. The answer is affirmative, as we will show in this work.
Apart from the Minkowski spacetime, other exact vacuum solutions have been found
for the action (\ref{action}) with no cosmological constant \cite{Salam:1976as}.
In this paper, we first generalize these solutions to include an arbitrary bare cosmological constant $\Lambda$, and then analyse in detail the dynamics of linear perturbations around a particularly interesting and simple configuration, in which the effect of the graviton mass and the cosmological constant terms compensate each other, and lead to a flat physical background. Around this configuration, a new gauge symmetry emerges in the quadratic action for perturbations, which removes the dynamical vector and scalar degrees of freedom, leaving the massless helicity-2 mode as the only dynamical state. Consequently, the dynamics of linearized fluctuations for
this theory around our flat background behaves {\it exactly} as in GR.

Our findings explicitly show that enhanced symmetry points exist in the space of background solutions of modified gravity models, such as massive gravity. These configurations might represent examples of consistent backgrounds where we can live on, and in which the infrared issues of GR could be addressed more successfully than in GR.

\smallskip

The paper is organized as follows. In section 2, we discuss   self-accelerating solutions in the generalised Fierz-Pauli model, including a bare cosmological constant. We  show that, imposing a particular tuning of parameters, the physical metric reduces to the Minkowski spacetime. In section 3, we study linear perturbations around this non-trivial flat space solution, using the ADM formalism. We split the analysis in tensor, vector and scalar modes with respect to the spatial metric and show that it only propagates the massless tensor modes as in GR. In section 4, we explicitly construct the gauge symmetry which reduces the number of physical degrees of freedom around our specific background, and we show how to express  it in a covariant way. Finally, we conclude in section 5 with a  discussion of our results.

\section{New exact solutions}

In order to study the theory (\ref{action}) and its exact solutions, it is convenient to implement the St\"uckelberg trick to recover diffeomorphism invariance following \cite{ArkaniHamed:2002sp}. This can be achieved by introducing a covariantization of the metric displacement $h_{\mu \nu}$ in terms
of the following definition
\be
g_{\mu \nu}\,=\,\eta_{\mu \nu}+h_{\mu \nu}\,=\,H_{\mu \nu}+\eta_{\alpha \beta}\,\partial_\mu \phi^\alpha \partial_\nu \phi^\beta
\,.
\ee
Therefore, the quantity $H_{\mu \nu}$ is given by a combination of the physical metric $g_{\mu\nu}$, and a metric in field space
\be\label{fpmetric}
f_{\mu \nu}\,\equiv\,\eta_{\alpha \beta}\,\partial_\mu \phi^\alpha \partial_\nu \phi^\beta\,.
\ee
The FP part of the Lagrangian density, in terms of the quantity $H_{\mu \nu}$, can be written as
 \be \label{fpl}
 {\cal L}_{FP}\,=\,-\frac{m^2}{4}\,\sqrt{-g}\,H_{\mu \nu} H_{\rho \sigma}\,\left(g^{\mu \rho} g^{\nu \sigma}-g^{\mu \nu } g^{\rho \sigma} \right)\,.
 \ee
The St\"uckelberg fields $\phi^{\alpha}$ have been introduced to restore reparametrisation invariance, so that the previous action is invariant under $x^{\alpha}\to x^\alpha+\xi^\alpha$, provided these St\"uckelberg fields $\phi^{\alpha}$ transform as scalars with respect to space-time symmetries (their indexes are raised and lowered by means of the fiducial metric $\eta_{\alpha \beta}$). Choosing the unitary gauge, defined by $\phi^\alpha=x^\alpha$, one obtains $H_{\mu \nu}\,=\,h_{\mu\nu}$ which leads to the theory written as in eq. (\ref{action}).  The simplest background solution for this theory, in absence of bare cosmological constant, is Minkowski space, given by
  \bea
  g_{\mu\nu}&=&\eta_{\mu\nu}
\hskip0.4cm,\hskip0.4cm
 \phi^\mu=x^\mu\hskip0.4cm,\hskip0.4cm H_{\mu \nu } = 0 \hskip0.4cm,\hskip0.4cm \Lambda=0\,.\label{soluno}
  \eea

Salam and Strathdee found other vacuum solutions, that we now generalize to the case of a non-zero
cosmological constant (see Appendix \ref{appA} for a full derivation). We write them in a coordinate system that is particularly useful for studying cosmology, where the metric reduces to the flat slicing of a maximally symmetric space, namely
\bea\label{metric}
d s^2\,=\,-d t^2+e^{2 \tilde{m} t}\,\left( d x_1^2+ d x_2^2+ d x_3^2\right)\, ,
\eea
while the St\"uckelberg fields acquire the following non-trivial background profile
\bea
\label{phi0}
\phi^{0}&=&
 \sqrt{\frac32}\, \frac{1}{\tilde{m}}\,\left[
 {\rm{arctanh}} {\left(\frac{2 \sinh{\tilde m t} +\tilde{m}^2\,r^2 \,e^{\tilde{m} t} }{2 \cosh{\tilde{m} t} -\tilde{m}^2\,r^2 \,e^{\tilde{m} t}  }\right)}
 -\tilde{m} \,r\,e^{\tilde{m} t} +
 {\rm{arctanh}} {\left( \tilde{m}\, r\,e^{\tilde{m} t}\right)}
 \right]\,,
 \\
%Sqrt[3/2] (-Sqrt[x1^2 + x2^2 + x3^2]  Exp[H0  t] +
%
  % )
\phi^{i}&=& \sqrt{\frac32}\,e^{\tilde{m} t}\,x^i \qquad (i=1,2,3)\,.
%\phi^{1}&=& \sqrt{\frac32}\,e^{\tilde{m} t}\,x^1 \,,\\
%\phi^{2}&=& \sqrt{\frac32}\,e^{\tilde m t}\,x^2 \,,\\
%\phi^{3}&=& \sqrt{\frac32}\,e^{\tilde m t}\,x^3 \,.
\eea
In the previous equations  we dubbed for simplicity $r\equiv \sqrt{x_1^2+x_2^2+x_3^2} $, and called
\be\label{mtilde}
\tilde m^2=\frac{m^2}{4}+\frac{\Lambda}{3}\,.
\ee
Notice that the Hubble scale is set by the FP mass and the cosmological constant, thus a
solution exhibits the acceleration if $\tilde m>0$. In absence of cosmological constant, $\Lambda=0$, and for a non-zero $m$, the solution exhibits the self-acceleration, i.e. an  acceleration without  cosmological constant.
%In Appendix A, we will prove that  this configuration solves the equations of motion associated with the action 
%(\ref{action}).

The general solution discussed above admits an interesting limit leading to flat space, when choosing the
cosmological constant  $\Lambda\,=\,-(3/4) \,m^2$. It is easy to check  that this background solution
for the metric and St\"uckelberg fields reads
%Notice that, this gen configuration allows to determine a version of Minkowski space {\bf GT rephrase/rethink
%to previous sentence. Discuss also the symmetries of the metric in field space}.
%Iin the limit $\Lambda=-(3/4) \,m^2$, one
 % recovers
%flat space in the metric, but not in the scalar fields, since in this limit
%\be
%\phi^{\mu}= \sqrt{\frac32}\, x^\mu\ee
%and the metric $g_{\mu\nu}\,=\,\eta$
 %instead of being
%$\phi^\mu\,=\, x^\mu $. In this limit, if $m\neq 0$, one has a 'strong coupling version' of flat space, again solution of the a massive
%theory.
%
% This means that this background is formally different
%from Minkowski space in the limit of vanishing FP mass. This property will be  crucial in our analysis of perturbations
% around this cosmological solution.
%
  \bea\label{soldue}
  g_{\mu\nu}&=&\eta_{\mu\nu}
\hskip0.4cm,\hskip0.4cm
 \phi^\mu=\sqrt{\frac32} x^\mu\hskip0.4cm,\hskip0.4cm H_{\mu \nu } = -\frac12 \,\eta_{\mu \nu} \hskip0.4cm,\hskip0.4cm \Lambda=-\frac34
  \,m^2\,.
  \eea
Notice that the previous solution is different from the simplest Minkowski solution in the absence of cosmological constant term (eqs.~(\ref{soluno})), by a constant in the St\"uckelberg fields.
%In the present context, the cosmological constant and the graviton mass are tuned in such a way to compensate each %other,
%and lead to a flat space solution with a non-trivial profile for the St\"uckelberg fields.
The field space metric $f_{\mu\nu}$, defined in
eq.~(\ref{fpmetric}), retains the same symmetries of the physical space-time metric $g_{\mu\nu}$.

In the next section, we will study the dynamics of perturbations around the background of eq.~(\ref{soldue}), showing that a new emerging gauge symmetry reduces the number of degrees of freedom to the ones of pure GR with no mass term and no cosmological constant.

%We have (indeces of $\phi^\mu$ are raised and lowered with flat Minkowski metric)

%\bea
%H_{\mu \nu}&=&-\frac12 \,\eta_{\mu \nu}+
%h_{\mu \nu}-\sqrt{\frac32}  \,\partial_{(\mu } \delta \phi_{\nu)}
%-\eta_{\alpha \beta} \,\partial_\mu \delta \phi^\alpha \partial_\nu \delta \phi^\beta\\
%&=&-\frac12 \,\eta_{\mu \nu}+\hat{H}_{\mu \nu}
%\eea
%where the $\hat H$ means perturbation. Notice
%that it is the first term in the right that characterizes the difference with respect to the 'normal branch' of Minkowski space.

% Plugging this expression in the FP action, we find
%\bea
%{\cal I}_{FP}&=&-\frac{m^2}{16}\,\int d^4 x   \sqrt{-^{(4)}g}\,\eta_{\mu \nu} \eta_{\rho \sigma}\,\left(g^{\mu \rho} g^{\nu \sigma}-g^{\mu \nu } g^{\rho \sigma} \right)\\&&
%+\frac{m^2}{4}\,\int d^4 x   \sqrt{-^{(4)}g}\,\eta_{\mu \nu} \hat{H}_{\rho \sigma}\,\left(g^{\mu \rho} g^{\nu \sigma}-g^{\mu \nu } g^{\rho \sigma} \right)\\
%&&-\frac{m^2}{4}\,\int d^4 x   \sqrt{-^{(4)}g}\,\hat{H}_{\mu \nu} \hat{H}_{\rho \sigma}\,\left(g^{\mu \rho} g^{\nu \sigma}-g^{\mu \nu } g^{\rho \sigma} \right)
%\eea

\section{Dynamics of perturbations}
\subsection{Definition of perturbations}

We follow an ADM approach and use Latin indexes to characterize quantities in the three spatial dimensions.
The metric reads
\be
d s^2\,=\,-N^2 \, d t^2+ \gamma_{i j } \,\left( d x^i + N^i d t\right)\, \left( d x^j + N^j d t\right)\,.
\ee
Perturbations to the lapse function and the shift vector are defined as
\bea
N&=& 1+ A\hskip0.5cm,\hskip0.5cm
N^i \,=\,{\cal B}^i\,,
\eea
and to the three dimensional metric as
\bea
\gamma_{i j}&=&\delta_{ij} +2 h_L \,\delta_{ij}+2\left( {\cal E}_{ij}-\frac13 \delta_{ij} {\cal E}_k^k \right)\,,
\eea
where the curvature perturbation is defined as ${\cal R}\,\equiv\,h_L-  \frac{1}{3}{\cal E}_k^k$.
Using the extrinsic curvature $K_{i j }\,\equiv\,\frac{1}{2 N}\left( \nabla_i N_{j} + \nabla_j N_{i}-\dot \gamma_{ij} \right)$, we express the Einstein-Hilbert part of the action (\ref{action}) in the usual form
\be
I_{EH}\,=\,\int d t d^3 x\,N \sqrt{\gamma}\,\left[
R-2 \Lambda + K_{\,i}^j K_{\,j}^i-K^2-2 \Lambda\right]\,.
\ee

\bigskip

Furthermore, it is convenient to decompose the various quantities into scalar, vector and tensors
with respect to the 3D spatial metric as
\bea
{\cal B}_i &=& S_i +B_{,i}\,,\\
{\cal E}_{ij}&=& h_{ij}+F_{(i,j)}+E_{, ij}\,,
\eea
with $S_i$ and $F_i$ transverse vectors, and $h_{ij}$ transverse traceless tensor, namely
\be
S^{i}_{,i} = F^{i}_{,i} = h^i_{\,i} = h^i_{\,j,\,i}\,=\,0\,.
\ee
The   perturbation of St\"uckelberg fields  split as
\be
\delta \phi_\mu\,=\,(\varphi,\,\chi_i+\psi_{,i})
\ee
with $\chi^i_{\,,i}=0$.

%\subsection{Action linear in perturbations} {\bf Do we really need this subsection? I think we can put something like ``The linear action of perturbations vanishes in agreement with (\ref{soldue}) being a solution of (\ref{action})''}

%The action (\ref{action})  expanded at linear level in perturbations around the background (\ref{soldue}) reads
%\bea
%I^{linear}&=&-\left( 2 \Lambda+\frac{3 m^2}{2}\right)\,\int dt d^3 x\,\left[
%A+3 {\cal R}
%\right]\,,
%\\
%&=&0
%\eea
%{\cal I}_{FP}^{lin}&=&-\frac{3 m^2}{2}\,\int dt d^3 x\,\left[
%A+3 {\cal R}
%\right]
%\eea
%while the Einstein-Hilbert part is
%\bea
%{\cal I}_{EH}^{lin}&=&-2 \Lambda\,\int dt d^3 x\,\left[
%A+3 {\cal R}\right]
%\eea
%once we choose  $\Lambda= -3/4 m^2$. This  is expected given that (\ref{soldue}) is a solution of equations of motion.

\subsection{Action quadratic in perturbations}

At  linear order, the  action of perturbations vanishes in agreement with (\ref{soldue}) being a solution of (\ref{action}). 
At quadratic order in perturbations around the background solution (\ref{soldue}), the EH contribution to the action (\ref{action}) reads
\bea
{I}_{EH}^{quad}\,=\,\int \,dt\, d^3x\,&&\Big\{  \dot{h}_{ij} \dot{h}^{ij}+ h^{ij} \Delta h_{ij}+2\Lambda h_{ij} h^{ij}
%\nonumber \\&&
+\frac12 \big(S^i -\dot{F}^i\big)^{,j}\big(S_i -\dot{F}_i\big)_{,j}\nonumber\\
&&-6 \dot{\cal R}^2+4 \dot{\cal R} \,\Delta(B-\dot{E})-4 A  \,\Delta {\cal R}+2 \,\partial_i
{\cal R} \partial^i {\cal R}\nonumber\\
%&&-2 {\cal R} \left(
%\Delta  {\cal E}_i^i-\nabla^i \partial^j {\cal E}_{ij}
%\right)
%\\
&&-2\Lambda \left(3 A {\cal R}+\frac32 {\cal R}^2+(A+{\cal R})\Delta E
+\frac12 (\Delta E)^2 -E_{,ij}E^{,ij}-F_{i,j} F^{i,j} \right)
\Big\}\, ,
\eea
where dots represent time derivatives. The FP part is given by
\bea
{ I}_{FP}^{quad}\,=\,-\frac{m^2}{4}\,\int \,dt \,d^3x\,&&\Big\{ \dot \varphi\,\left[
3 \sqrt{6} A+ 3 \sqrt{6}  \, \Delta E+ 9 \sqrt{6} {\cal R} -6 \,\Delta  \psi -3 \dot \varphi \right] \nonumber\\
&& + A\,\left[
-\frac92 A- 3  \,\Delta  E- 9  {\cal R} +3 \sqrt{6}\,\Delta   \psi  \right] \nonumber \\
&&+{\cal R}\,\left[
- 21\,\Delta   E- \frac{63}{2}  {\cal R} +9 \sqrt{6} \,\Delta\psi  \right] \nonumber \\
&&-3 \left(\Delta  \psi  -\sqrt{\frac{ {3} }{ 2}}\Delta E \right)^2-6 h_{ij} h^{ij}-6 F_{i,j} F^{i,j}
-6 E_{,ij}E^{,ij}+3 (\Delta E)^2 \Big\}\,.
\eea
Adding these two pieces, and setting $\Lambda= -3/4 m^2$, we obtain that the tensor and vector contributions organize in such a way to  become {\it exactly} the ones of pure GR in the Minkowski background, namely
\bea\label{tenact}
{ I}_{tens, vect}^{quad}&=&\int dt d^3x\,\Big[  \dot{h}_{ij} \dot{h}^{ij}+ h^{ij} \Delta h_{ij}+\frac12 \left(S^i -\dot{F}^i\right)^{,j}\left(S_i -\dot{F}_i\right)_{,j}\Big]\,.
\eea
Therefore, they behave as in pure GR, describing the propagation of helicity-2 states.
One can prove this standard fact by a Hamiltonian analysis of the action (\ref{tenact}), as  shown in detail in Appendix \ref{appB}.

After fixing $\Lambda= -3/4 m^2$, the Lagrangian density associated with the total scalar contribution, up to a total derivative, is
\bea
{\cal L}^{quad}_{scal}\,=\,&&\Big\{
-6 \dot{\cal R}^2+4 \dot{\cal R} \,\Delta(B-\dot{E})-4 A \, \,\Delta {\cal R}-2 {\cal R} \Delta {\cal R}\Big\} \nonumber\\
%&&-2 {\cal R} \left(
%\Delta  {\cal E}_i^i-\nabla^i \partial^j {\cal E}_{ij}
%\right)
%\\
%&&+\frac{3 m^2}{2} \left(3 A {\cal R}+\frac32 {\cal R}^2+(A+{\cal R})\Delta E
%+\frac12 (\Delta E)^2\right)
%\Big\}\\
&&\nonumber-\frac{m^2}{4}\,
\Big\{
\dot \varphi\,\left[
3 \sqrt{6} A+ 3 \sqrt{6}  \, \Delta E+ 9 \sqrt{6} {\cal R} -6 \,\Delta  \psi -3 \dot \varphi \right]\\
&&\nonumber + A\,\left[
-\frac92 A- 9  \,\Delta  E- 27  {\cal R} +3 \sqrt{6}\,\Delta   \psi  \right]\\
&&+{\cal R}\,\left[
- 27\,\Delta   E- \frac{81}{2}  {\cal R} +9 \sqrt{6} \,\Delta\psi  \right]\,
-3 \left(\Delta  \psi  -\sqrt{\frac{ {3} }{ 2}}\Delta E \right)^2 \Big\}\label{lagscal}\,.
\eea

\subsection{Hamiltonian analysis of the scalar sector}
It is convenient to perform a Hamiltonian analysis of this system, in order to count the number of dynamical degrees of freedom in the scalar sector described by the quadratic Lagrangian (\ref{lagscal}). We first notice that this Lagrangian does not contain time derivatives of $A$, $B$, and $\psi$; these quantities will then be associated with constraints. Therefore, we have in principle three dynamical scalar degrees of freedom: ${\cal R}$, ${E}$ and $\varphi$. The conjugate momenta associated with these dynamical variables are
 \bea
 \Pi^{\cal R}&=& \frac{\partial{\cal L}_{s}}{\partial\dot{\cal R}}\ =\  4 \Delta B-4 \Delta \dot{E}-12 \dot{{\cal R}} \,,\\
 \Pi^{E}&=& \frac{\partial{\cal L}_{s}}{\partial\dot{\cal E}}\ =\ -4 \Delta \dot{{\cal R}} \,,\\
 \Pi^{\varphi}&=&\frac{\partial{\cal L}_{s}}{\partial\dot{\varphi }}\ =\ -\frac34 m^2 \left(\sqrt{6} A +\sqrt{6} \Delta E-2\dot{\varphi}-2 \Delta \psi +3 \sqrt{6}
 {\cal R}
 \right)\,.
 \eea
The scalar Hamiltonian is   defined as
\be
{\cal H}_{s}\,=\,\dot{\cal R} \,\Pi^{\cal R}+ \dot{E} \,\Pi^{\cal E}
+\dot{\varphi} \,\Pi^{\varphi}-{\cal L}_s+A \,{\cal C}_A+B\, {\cal C}_B +\psi \,{\cal C}_\psi
\ee
where we introduce the constraint ${\cal C}_A\,=\,\partial{\cal L}_{s}/{\partial{ A}}$, and similarly for $B$ and $\psi$, so that ${\cal H}_{s}$ is
   written in the following way
%By means of the conjugate momenta, the Hamiltonian  for our system can be written as
\bea
{\cal H}_s&=&\frac{1}{24}\Big[ 9 \,\left(\Delta^{-1} {\Pi}^E \right)^2
%+\frac{16}{m^2}\,\left({\Pi}^\varphi\right)^2
-6\,\left(\Delta^{-1}\Pi^E\right)\,\Pi^{\cal R}
+4\,\Pi^\varphi\,\left( \frac{2\,\Pi^\varphi }{m^2} +  9 \sqrt{6}\,{\cal R}  \right)
+12\,\Delta\left( \sqrt{6} E \Pi^\varphi+4{\cal R}^2\right)
%++
%\left(\Delta^{-1}\Pi^E\right)\,\left( 9 \sqrt{6}\, m^2 \,\varphi -4\, \Pi^{\cal R}\right)\nonumber\\
%&&+486\,m^2\,{\cal R}^2+12 \,\Delta\,\left(2 \sqrt{6}\,\Pi^\varphi\,E-27 \,m^2\,E \,{\cal R}
%+8\,{\cal R}^2\right)
\Big].
\eea
The constraints read
\bea
{\cal C}_{A}&=&-4 \Delta {\cal R}-\sqrt{\frac32} \Pi^\varphi
\,,\\
{\cal C}_{B}&=&-\Pi^E\,,\\
{\cal C}_\psi&=& \Delta \Pi^\varphi\,.
\eea
These   are first class constraints, since their mutual
Poisson brackets vanish: 
\be
\left\{ {\cal C}_A,\,{\cal C}_B\right\}\,=\,\left\{ {\cal C}_A,\,{\cal C}_\psi\right\}
\,=\,\left\{ {\cal C}_\psi,\,{\cal C}_B\right\}\,=\, 0\,.
\ee
Moreover, the Poisson brackets among the  constraints and the Hamiltonian satisfy the following relations
\bea
\left\{ {\cal C}_A,\,{\cal H}_s\right\}&=&-{\cal C}_B\,,\\
\left\{ {\cal C}_\psi,\,{\cal H}_s\right\}
&=& 0\,,\\
\left\{ {\cal C}_B,\,{\cal H}_s\right\}
&=&\sqrt{\frac32}\,{\cal C}_\psi\,,
\eea
which imply that the constraints are preserved under time evolution.
After imposing these constraints, it is straightforward
to check that the Hamiltonian vanishes. Indeed three first class constraints are able to remove all the
phase-space dynamical degrees of freedom.
Consequently, also the scalar sector does not contain any physical degree of freedom.

\smallskip

To summarize, quadratic perturbations around the flat-space background (\ref{soldue}) lead to the propagation
of only helicity-2 modes, exactly as in GR.

\subsection{The new gauge symmetry}
We can understand the result of the previous section in terms of gauge symmetries holding for the action of
  quadratic perturbations around the background solution (\ref{soldue}). Starting from our Hamiltonian analysis,
we can implement the standard rules for obtaining the gauge symmetry associated with the first class constraints (see for example \cite{Mukhanov:1994zn}).  The infinitesimal transformations that leave the quadratic scalar Lagrangian  (\ref{lagscal}) invariant,  up to total derivatives, 
read
\bea
\delta \varphi&=&\sqrt{\frac32}\epsilon_A-\Delta \epsilon_\psi\,,
\label{gau1}
\\
\delta \psi&=&\dot \epsilon_\psi+\sqrt{\frac32}\epsilon_B \,,\label{gau5}\\
\delta {\cal R}&=&0\label{gau2}\,,\\
\delta E&=&\epsilon_B\label{gau3}\,,\\
\delta A&=&\dot \epsilon_A\label{gau4}\,,\\
\delta B&=& \dot\epsilon_B-\epsilon_A\label{gau6}\,,
\eea
for three arbitrary functions $\epsilon_A$, $\epsilon_B$ and $\epsilon_\psi$.
The parameters $\epsilon_A$ and $\epsilon_B$ are associated with the
diffeomorphism invariance of the FP action, once the St\"uckelberg fields are introduced. The symmetry associated with
$\epsilon_\psi$, that acts only on the St\"uckelberg scalars, is the new symmetry.

\smallskip

It is not difficult to express these  symmetries in a covariant way.  Adding the cosmological constant 
 contribution to 
the covariant FP action (\ref{fpl}), tuning $\Lambda=-3/4 \,m^2$, and expanding at quadratic order in perturbations
around our solution (\ref{soldue}), one finds the following covariant Lagrangian density
% added to the cosmologo
%
 %in a covariant way. Only the terms involving the scalar
%perturbations will be important, since as we mentioned the symmetry only acts on them. After some manipulation one finds the following contributions to the FP Lagrangian 
\be\label{resquad}
{\cal L}^{FP+\Lambda}_{quad}\,=\, \frac{9}{32} m^2 \left( h_{\mu}^{\mu} -\sqrt{\frac{8}{3}} \partial_{\mu} \phi^{\mu} \right)^2.
\ee
It is easy to check that this Lagrangian is invariant, at quadratic order, under the transformation
\bea\label{newsym}
 \delta \phi^\mu &=& \delta \phi^\mu - \sqrt{\frac{3}{2}} \xi^{\mu}+\chi^\mu\hskip0.7cm {\text{with}}\hskip0.7cm\partial_\mu \chi^\mu\,=\,0,  \\
 h_{\mu \nu} &=& h_{\mu \nu} - \xi_{(\mu,\nu)}\,.
 \eea
Focussing on  scalar perturbations, the symmetry (\ref{newsym})
corresponds exactly to the extra symmetry parameterized by the function $\epsilon_\psi$ in eqs (\ref{gau1}) and (\ref{gau5}). The existence of this gauge symmetry, besides the standard diffeomorphism invariance, implies that the scalar and vector degrees of freedom do not propagate around the background (\ref{soldue}). Note that this resulting quadratic action (\ref{resquad})
 is exactly the same as a graviton mass term proportional to $(H^{\mu}_{\,\mu})^2$. It is known that this mass term does not modify GR, i.e. it leads  only to a  massless transverse-traceless graviton \cite{Rubakov}, which is consistent with our result.  On a different
 setting,  it was also shown in   \cite{Deser:2001wx} that, thanks to a new gauge symmetry,  massive gravity on  a de Sitter background
   does not propagate scalar zero modes, when the size of the cosmological constant is tuned to a precise value  depending on the graviton mass.

\section{Discussion}

We constructed new solutions for a covariantized non-linear Fierz-Pauli theory of massive gravity, equipped with a cosmological constant $\Lambda$. We focused on a particularly simple configuration corresponding to Minkowski space with a non-trivial profile for the St\"uckelberg fields. The space-time flatness is achieved by tuning $\Lambda$ to a particular value related to the graviton mass. We then studied linear perturbations around this  configuration,
and showed that it only propagates the massless helicity-2 modes, in the same way as pure GR without a cosmological constant.
We interpreted this behavior as due to a new gauge symmetry acting on the St\"uckelberg fields, in addition to  diffeomorphism invariance.

% corresponding to Minkowski space with a
%non-trivial  profile for the St\"uckelberg fields. 

% in the simplest non-linear extension of FP with a bare cosmological constant $\Lambda$, we found a new gauge symmetry for the St\"uckelberg field. The background solution corresponds to a flat metric with a non-trivial profile for the St\"uckelberg fields, where the spacetime flatness is achieved by a particular choice of $\Lambda$. The new group of invariance is large enough to elminate all the vector and scalar perturbations.
%In our case, the resulting theory propagates only the massless helicity-2 modes, in the same way as pure GR without a cosmological constant.

It would be important to understand whether this new symmetry is only  associated with  the particular 
set-up discussed here, or whether it arises more generally. In other non-linear massive gravity models, several examples of self-accelerating solutions have been found \cite{us1, us2, us3, D'Amico:2011jj, Gumrukcuoglu:2011ew, shinji,  pilo}. Recently, \cite{shinji}  found similar situations where there are  no propagating scalar and vector modes around self-accelerating open-FRW solutions in the non-linear massive gravity models proposed by de Rham, Gabadadze and Tolley \cite{deRham:2010kj}. See also \cite{pilo} for examples of self-accelerating
solutions in non-linear theories  of massive gravity, around which there are no additional degrees of freedom besides tensor modes. 
 One one hand, a common feature between our solution and the set-up of \cite{shinji} is that the metric in field space preserves the same symmetries as the physical metric, which in our case correspond to the fact that both are Minkowski spacetime. On the other hand, there are other self-accelerating de Sitter solutions in the same theory for which the field space metric is not invariant under same isometries as the physical metric \cite{us1, us2, us3}. Those solutions are indeed similar to the ones obtained in this article, but where the cosmological constant is not tuned with the mass term to give flat space. The non-trivial profiles for the St\"uckelberg fields render the analysis of the dynamics of fluctuations more difficult. However, it would be interesting to study whether new symmetries also arise while considering perturbations around these backgrounds.

Finally, it is also important to explore whether the new gauge symmetry holds only at linear order in perturbations, or survive at higher orders. If the latter is the case, then these configurations may represent examples of consistent backgrounds where we can live on, and in which the infrared issues of GR could be addressed more successfully than in GR. For example, one could envisage a mechanism that dynamically tunes the graviton mass with the cosmological constant, leading to flat space also in the presence of $\Lambda$. We will investigate these interesting questions elsewhere.

%{\bf GN maybe you want to include more ideas in order to include all the points:}
%
%\begin{itemize}
%\item Say that it might be manifestation of strong coupling behavior: on the other hand the stable vacuum with enhanced
%symmetry exists, and it might be that we're living there.
%\item It's straightforwardly extended to non-linear realization of massive gravity, for which analogous configurations exist \cite{}.
%\item Understand the rule of the structure of the field profile (metric in field space).
%\item There's a tuning to do to relate cc with mass: would be great to find a dynamical way to ensure it, as coupling with scalar field
%as realized in some set-ups in \cite{}. Generalize to the case in which the tuning is not imposed.
%\item On the other hand, it might be that it's just GR written in a special way. Answers to all these questions will be discussed elsewhere. {\bf GN cannot be the case case in shinji's solution, so i don't think we should say it}
%\end{itemize}

\subsection*{Acknowledgments}
GT is supported by an STFC Advanced Fellowship ST/H005498/1. KK is supported by supported by STFC grant ST/H002774/1, ERC and the Leverhulme trust. 

\bigskip
%\mbox{}\\[-2.9em]
\begin{appendix}

\section{Derivation of the exact solutions}\label{appA}
We begin by choosing the gauge $\phi^\mu=x^\mu$, which implies that $f_{\mu\nu}=\eta_{\mu\nu}$. Since we are interested in static spherical solutions to the action (\ref{action}), we introduce the ansatz
\be\label{ansmetr}
d s^2\,=\,-C(r) \,d t^2+A(r)\, d r^2 +2 D(r)\, dt dr+B(r) d \Omega^2,
\ee
where $d \Omega^2 = d \theta^2 + \sin^2 \theta d \phi^2$, and to simplify expressions, one can further choose to write the field space metric $f_{\mu\nu}$ in spherical coordinates as $f_{\mu\nu}dx^\mu dx^\nu=-dt^2 + dr^2 + r^2 d\Omega^2$.

After introducing the ansatz (\ref{ansmetr}) in the equations of motion derived from the action (\ref{action}), one can show there is a constraint which leads to two branches of solutions: one with $D(r)=0$ and the other where $B(r)=3 r^2/2$. The first branch was studied in \cite{Aragone:1972fn}, but here, we are interested in the second branch, where exact solutions were initially found by Salam and Strathdee \cite{Salam:1976as}. In this second class, a new constraint enforces $\Delta(r)  \,=\,A(r) C(r)+D^2(r) \,\equiv\,\Delta_0\,=\,\mathrm{const}$, and the rest of equations of motion admit the following solution
\bea\label{solmetr}
A(r) &=& \frac{3 \Delta_0}{2} (p(r) + \alpha +1),\qquad
B(r) =\frac{3}{2} r^2,\\
C(r) &=&\frac{3 \Delta_0}{2}(1 - p(r)), \qquad
D(r) = \frac{3}{2} \Delta_0 \sqrt{p(r)(p(r)+\alpha)}, \nonumber
\eea
where 
\be
p(r)=\frac{c}{r} + \frac{2\tilde{m}^2 r^2}{3 \Delta_0^{3/2}}, \qquad
\alpha =\frac{4}{9\,\Delta_0}-1,
\ee
and $\tilde{m}$ is defined as in (\ref{mtilde}). $\Delta_0$ and $c$ are two integration constants that obey $c \geq 0$ and $0<\sqrt{\Delta_0} \leq 2/3$ for $D(r)$ to be real. In this work we only consider $c=0$ and the extremal value $\Delta_0=4/9$. For these values, the solution (\ref{solmetr}) can be re-casted in the simple FRW form of eq.~(\ref{metric}) by the following coordinate transformation
\bea
x^0\equiv t&\rightarrow& \sqrt{\frac32}\, \frac{1}{\tilde{m}}\,\left[
 {\rm{arctanh}} {\left(\frac{2 \sinh{\tilde m t} +\tilde{m}^2\,r^2 \,e^{\tilde{m} t} }{2 \cosh{\tilde{m} t} -\tilde{m}^2\,r^2 \,e^{\tilde{m} t}  }\right)}
 -\tilde{m} \,r\,e^{\tilde{m} t} +
 {\rm{arctanh}} {\left( \tilde{m}\, r\,e^{\tilde{m} t}\right)}
 \right]\,,\nonumber
 \\
x^{i}&\rightarrow& \sqrt{\frac32}\,e^{\tilde{m} t}\,x^i \qquad (i=1,2,3)\,.
\eea
The last transformation implies that $r \rightarrow \sqrt{\frac32}\,e^{\tilde{m} t}\,r$, while $\theta$ and $\phi$ remain unchanged. Furthermore, the transformations change our initial gauge so that $f_{\mu\nu}\neq \eta_{\mu\nu}$, since the initial St\"uckelberg fields $\phi^\mu=x^\mu$ get modified  to those in (\ref{phi0}).

\section{Hamiltonian analysis of tensor and vector degrees of freedom}
\label{appB}
The canonical momenta associated to the tensor and vector modes in the action (\ref{tenact}) are
\bea
\Pi^{ij}_h&=&\frac{\delta {\cal L}^{quad}_{tens,vect}}{\delta \dot{h}_{ij}}\ =\ 2\dot{h}^{ij} \nonumber \\
\Pi^{i}_F&=&\frac{\delta {\cal L}^{quad}_{tens,vect}}{\delta \dot{F}_{i}}\ =\ \Delta(S^i-\dot{F}^{i}),
\eea
where ${\cal L}^{quad}_{tens,vect}$ is associated Lagrangian density of the action (\ref{tenact}).
In momentum's language, the Lagrangian ${\cal L}^{quad}_{tens,vect}$ then reads
\bea
{I}_{tens, vect}^{quad}&=&\int dt d^3x\,\Big[  \Pi_{h}^{ij} \dot{h}_{ij}+\Pi_{F}^{i} \dot{F}_{i}-{\cal H}_{t}-{\cal H}_v-S_i\Pi^i \Big]\, ,
\eea
where
\be
{\cal H}_t=\frac{1}{4}\Pi_{h}^{ij} \Pi^{h}_{ij}-{h}^{ij} \Delta {h}_{ij},\qquad \qquad {\cal H}_v=-\frac{1}{2}\Pi^{F}_{i}\Delta^{-1} \Pi_{F}^{i},
\ee
with $\Delta^{-1}$ the inverse Laplace operator. Let us first consider the tensor modes, which do not have any associated constraint. Therefore, both tensor modes are physical degrees of freedom, and their Hamiltonian is exactly that of GR (see for example \cite{Gong:2010xp}). In the vector case, $F^i$ is the only dynamical degree, while $S^i$ appears as a Lagrange multiplier, enforcing the constraints $\Pi^i_F=0$. These constraints are first class since they commute among each other and with the Hamiltonian ${\cal H}_v$, thus they represent the two gauge modes associated with vector perturbations. In summary, the tensor modes as the same as in GR and there are no physical vector degrees of freedom.

\end{appendix}

\bigskip
%\mbox{}\\[-2.9em]


\footnotesize
\begin{thebibliography}{nn}
\small

%\cite{Fierz:1939ix}
\bibitem{Fierz:1939ix}
  M.~Fierz and W.~Pauli,
  %``On relativistic wave equations for particles of arbitrary spin in an electromagnetic field,''
  Proc.\ Roy.\ Soc.\ Lond.\ A {\bf 173} (1939) 211.
  %%CITATION = PRSLA,A173,211;%%

%\cite{Boulware:1973my}
\bibitem{Boulware:1973my}
  D.~G.~Boulware and S.~Deser,
  %``Can gravitation have a finite range?,''
  Phys.\ Rev.\ D {\bf 6} (1972) 3368.
  %%CITATION = PHRVA,D6,3368;%%



  %\cite{Hinterbichler:2011tt}
\bibitem{Hinterbichler:2011tt}
  K.~Hinterbichler,
  %``Theoretical Aspects of Massive Gravity,''
  arXiv:1105.3735 [hep-th].
  %%CITATION = ARXIV:1105.3735;%%



%\cite{Salam:1976as}
\bibitem{Salam:1976as}
  A.~Salam and J.~A.~Strathdee,
  %``A Class of Solutions for the Strong Gravity Equations,''
  Phys.\ Rev.\ D {\bf 16} (1977) 2668.
  %%CITATION = PHRVA,D16,2668;%%

%\cite{ArkaniHamed:2002sp}
\bibitem{ArkaniHamed:2002sp}
  N.~Arkani-Hamed, H.~Georgi and M.~D.~Schwartz,
  %``Effective field theory for massive gravitons and gravity in theory space,''
  Annals Phys.\  {\bf 305} (2003) 96
  [hep-th/0210184].
  %%CITATION = HEP-TH/0210184;%%

%%\cite{Cicoli:2012tz}
%\bibitem{Cicoli:2012tz}
 % M.~Cicoli, F.~G.~Pedro and G.~Tasinato,
 % %``Natural Quintessence in String Theory,''
 % arXiv:1203.6655 [hep-th].
 % %%CITATION = ARXIV:1203.6655;%%

%\cite{Mukhanov:1994zn}
\bibitem{Mukhanov:1994zn}
  V.~F.~Mukhanov and A.~Wipf,
  %``On the symmetries of Hamiltonian systems,''
  Int.\ J.\ Mod.\ Phys.\ A {\bf 10} (1995) 579
  [hep-th/9401083].
  %%CITATION = HEP-TH/9401083;%%
  
%\cite{Rubakov:2008nh}
\bibitem{Rubakov}
  V.~A.~Rubakov and P.~G.~Tinyakov,
  %``Infrared-modified gravities and massive gravitons,'' 
  Phys.\ Usp.\  {\bf 51}, 759 (2008)  [arXiv:0802.4379 [hep-th]].  %%CITATION = ARXIV:0802.4379;%%
  
  %\cite{Deser:2001wx}
\bibitem{Deser:2001wx}
  S.~Deser and A.~Waldron,
  %``Stability of massive cosmological gravitons,''
  Phys.\ Lett.\ B {\bf 508} (2001) 347
  [hep-th/0103255].
  %%CITATION = HEP-TH/0103255;%%

   %\cite{Koyama:2011xz}
\bibitem{us1}
  K.~Koyama, G.~Niz and G.~Tasinato,
  %``Analytic solutions in non-linear massive gravity,''
  Phys.\ Rev.\ Lett.\  {\bf 107} (2011) 131101
  [arXiv:1103.4708 [hep-th]].
  %%CITATION = ARXIV:1103.4708;%%

%\cite{Koyama:2011yg}
\bibitem{us2}
  K.~Koyama, G.~Niz and G.~Tasinato,
  %``Strong interactions and exact solutions in non-linear massive gravity,''
  Phys.\ Rev.\ D {\bf 84} (2011) 064033
  [arXiv:1104.2143 [hep-th]].
  %%CITATION = ARXIV:1104.2143;%%


%\cite{Koyama:2011wx}
\bibitem{us3}
  K.~Koyama, G.~Niz and G.~Tasinato,
  %``The Self-Accelerating Universe with Vectors in Massive Gravity,''
  JHEP {\bf 1112} (2011) 065
  [arXiv:1110.2618 [hep-th]].
  %%CITATION = ARXIV:1110.2618;%%
  
   
%\cite{D'Amico:2011jj}
\bibitem{D'Amico:2011jj}
  G.~D'Amico, C.~de Rham, S.~Dubovsky, G.~Gabadadze, D.~Pirtskhalava and A.~J.~Tolley,
  %``Massive Cosmologies,''  
  Phys.\ Rev.\ D {\bf 84}, 124046 (2011)  [arXiv:1108.5231 [hep-th]].  %%CITATION = ARXIV:1108.5231;%%


%\cite{Gumrukcuoglu:2011ew}
\bibitem{Gumrukcuoglu:2011ew}
  A.~E.~Gumrukcuoglu, C.~Lin and S.~Mukohyama,
  %``Open FRW universes and self-acceleration from nonlinear massive gravity,''  
  JCAP {\bf 1111}, 030 (2011)  [arXiv:1109.3845 [hep-th]].  %%CITATION = ARXIV:1109.3845;%%

%\cite{Gumrukcuoglu:2011zh}
\bibitem{shinji}
  A.~E.~Gumrukcuoglu, C.~Lin and S.~Mukohyama,
  %``Cosmological perturbations of self-accelerating universe in nonlinear massive gravity,''
  JCAP {\bf 1203} (2012) 006
  [arXiv:1111.4107 [hep-th]].
  %%CITATION = ARXIV:1111.4107;%%



%\cite{Crisostomi:2012db}
\bibitem{pilo}
  M.~Crisostomi, D.~Comelli and L.~Pilo,
  %``Perturbations in Massive Gravity Cosmology,''
  arXiv:1202.1986 [hep-th].
  %%CITATION = ARXIV:1202.1986;%%


  %\cite{deRham:2010kj}
\bibitem{deRham:2010kj}
  C.~de Rham, G.~Gabadadze and A.~J.~Tolley,
  %``Resummation of Massive Gravity,'' 
   Phys.\ Rev.\ Lett.\  {\bf 106}, 231101 (2011)  [arXiv:1011.1232 [hep-th]].  %%CITATION = ARXIV:1011.1232
  


%\cite{Aragone:1972fn}
\bibitem{Aragone:1972fn} 
  C.~Aragone and J.~Chela-Flores,
  %``Properties of the f-g theory,''
  Nuovo Cim.\ A {\bf 10}, 818 (1972).
  %%CITATION = NUCIA,A10,818;%%


%\cite{Gong:2010xp}
\bibitem{Gong:2010xp}
  J.~-O.~Gong, S.~Koh and M.~Sasaki,
  %``A complete analysis of linear cosmological perturbations in Ho\v{r}ava-Lifshitz gravity,''
  Phys.\ Rev.\ D {\bf 81} (2010) 084053
  [arXiv:1002.1429 [hep-th]].
  %%CITATION = ARXIV:1002.1429;%%




\end{thebibliography}
\end{document}